\documentclass[preprint,showpacs,preprintnumbers,amsmath,amssymb,endfloats*]{revtex4}
\usepackage{graphicx}

\begin{document}
\def \ee {\varepsilon}
\thispagestyle{empty}
\title{
Comment on ``Anomalies in electrostatic calibration for the
measurement of the Casimir force in a sphere-plane geometry''
}

\author{R.~S.~Decca,${}^1$ E.~Fischbach,${}^2$
G.~L.~Klimchitskaya,${}^3$\footnote{on leave from
North-West Technical University, St.Petersburg, Russia}
D.~E.~Krause,${}^{4,2}$ D.~L\'{o}pez,${}^5$
U.~Mohideen,${}^6$
and V.~M.~Mostepanenko${}^3$\footnote{on leave from Noncommercial Partnership
``Scientific Instruments'',  Moscow,  Russia}
}

\affiliation{
${}^1$Department of Physics, Indiana University-Purdue
University Indianapolis, Indianapolis, Indiana 46202, USA\\
${}^2$Department of Physics, Purdue University, West Lafayette, Indiana
47907, USA\\
${}^3$Center of Theoretical Studies and Institute for Theoretical
Physics, Leipzig University,
D-04009, Leipzig, Germany \\
${}^4$Physics Department, Wabash College, Crawfordsville, Indiana 47933,
USA\\
${}^5$Center for Nanoscale Materials, Argonne National Laboratory,
Argonne, Illinois 60439, USA \\
${}^6$Department of Physics and Astronomy, University of California,
Riverside, California 92521, USA
}

\begin{abstract}
Recently W. J. Kim, M. Brown-Hayes, D. A. R. Dalvit, J. H. Brownell,
and R.\ Onofrio [Phys. Rev. A {\bf 78}, 036102(R) (2008)] performed
electrostatic calibrations for a plane plate above a
centimeter-size spherical lens at separations down to 20-30\,nm and 
observed ``anomalous behavior''. It was found that the gradient of 
the electrostatic force does not depend on separation as
predicted on the basis of a pure Coulombian contribution.  Some
hypotheses which could potentially explain the deviation from the 
expected behavior were considered, and qualitative arguments in 
favor of the influence of patch surface potentials were presented. 
We demonstrate that for the large lenses at separations of a few 
tens nanometers from the plate, the electrostatic force law used by 
the authors is not applicable due to possible deviations of the 
mechanically polished and ground lens surface from a perfect 
spherical shape. A model is proposed which explains the observed 
``anomalous behavior'' using the standard Coulombian force.
\end{abstract}

\maketitle

In Ref.~\cite{Onofrio}, anomalies in the electrostatic calibration for the
measurement of the Casimir force in a sphere-plane geometry were found.
Precision electrostatic calibrations in the sphere-plane geometry have
attracted much attention in the last few years in connection
with measurements of the Casimir force 
\cite{2,3,4,5,6,7,8,9,10,11,12,13,14,15,16}.
In these measurements electrostatic calibrations play an important role. 
They allow precise independent determination of such basic
quantities as absolute separation, cantilever spring constants,
sphere radii, parameters of the micromechanical oscillator, and the contact
potential difference of the grounded test bodies. Because of this, any
inaccuracy in the theoretical expression for the electric force used in
the calibration introduces additional systematic errors in the measurement
data for the Casimir force and invites questions on the validity of the
experimental results that are obtained.

Reference \cite{Onofrio} presents the experimental data from electrostatic 
calibrations in the configuration of a Si plate above a large spherical
lens of radius $R=30.9\pm 0.15\,$mm, both covered with an Au film.
In these calibrations, separation distances $d$ down to a few tens of
nanometers from the point of contact between the plate and the sphere
were explored. Surprisingly, instead of the expected $d^{-2}$ distance
dependence of the gradient of the electric force, as is given by the
main contribution to the exact result in the sphere-plate configuration 
\cite{11} or,
equivalently, by the proximity force approximation, a dependence of
order $d^{-1.7}$ was observed from four separate experimental sequences.
The values of the contact potential difference $V_c$, in at least
two sequences, were found to be separation-dependent. 
Reference \cite{Onofrio} discusses five hypotheses which could potentially 
explain a deviation from the expected force law, specifically, static
deflection of the cantilever, thermal drift, nonlinearity of the
piezoelectric transducer, nonlinear oscillations of the cantilever, and the
surface roughness. It was found that none of these explain the
anomaly. A sixth
hypothesis, favored by the authors, is the effect of patch surface
potentials. However, no specific arguments in its favor were provided, 
except for the
observation that $V_c$ is separation-dependent in at least two
sequences. This is, however, simply an observation 
that the electric force gradient behaves anomalously, rather than a
determination of the specific physical cause. On this basis the
authors argue that their ``findings affect the accuracy of the electrostatic
calibrations and invite reanalysis of previous determinations of the 
Casimir force''.

Below we demonstrate that the observed anomalies find a clear explanation 
using the standard distance-dependence of the electric force, if one takes 
into account deviations of the lens surface from a perfect spherical shape.
Such deviations are unavoidably present on any spherical surface of
centimeter size. Hence, they preclude the use of the
simplest formulation of the proximity
force approximation for a constant radius of curvature at short
separations as used in the paper. In the
conclusion we formulate some basic requirements for
precision calibration procedures, and 
emphasize that all previous experiments
on the measurement of the Casimir force 
\cite{2,3,4,5,6,7,8,9,10,11,12,13,14,15,16}
are absolutely irrelevant to the phenomenon observed in \cite{Onofrio}
because they are performed at large separations \cite{2}, or with
spheres of much smaller radii \cite{3,4,5,6,7,8,9,10,11,12,13,14,15,16}.

Using the proximity force approximation \cite{Blocki}, paper \cite{Onofrio}
represents the gradient of the electric force between 
a centimeter-size spherical
lens and plate as
\begin{equation}
F_{\rm el}^{\prime}=\pi\epsilon_0\frac{R(V-V_c)^2}{d^2},
\label{eq1}
\end{equation}
\noindent
where  $V$ is the applied voltage,
$V_c$ is the contact potential, $d$ is the gap separation, and
$\epsilon_0$ is the permittivity of vacuum
(the minus sign on the right-hand side of this formula in \cite{Onofrio}
is a misprint).
The frequency shift of the cantilever due to an
external force is given by
\begin{equation}
\nu^2-\nu_0^2=-\frac{1}{4\pi^2m_{\rm eff}}\,F_{\rm el}^{\prime},
\label{eq2}
\end{equation}
\noindent
where $m_{\rm eff}$ is the effective mass of the oscillator.
Using Eq.~(\ref{eq1}),
this frequency shift can be rearranged to the form
\begin{equation}
\nu^2-\nu_0^2=-k_{\rm el}(d)\,(V-V_0)^2, \qquad
k_{\rm el}(d)=\frac{\epsilon_0R}{4\pi m_{\rm eff}d^2}.
\label{eq3}
\end{equation}
\noindent
However, as noted in \cite{Onofrio}, the experimental data from four
separate sequences follow a power law, similar to the $d^{-2}$ dependence
in Eqs.~(\ref{eq1}), (\ref{eq3}), but with powers $-1.70\pm 0.01$,
$-1.77\pm 0.02$, $-1.80\pm 0.01$, and $-1.54\pm 0.02$, far from the
expected value of $-2$.

As mentioned above,
Ref.~\cite{Onofrio} discusses several hypotheses which could explain 
the observed anomaly and discards all of them. As a possible explanation  the
effect of patch surface potentials was considered, but only qualitative
arguments that this effect might be responsible for the observed
anomalous behavior of
the electrostatic force were provided. These arguments, however,
do not take into account Refs.~\cite{8,10}, where the role
of patches due to the grains of polycrystalline metal film in the
measurements of the Casimir force by means an atomic force microscope
\cite{8} and a micromechanical torsional oscillator \cite{10}
was specifically investigated in detail. Thus, in \cite{8} it was concluded
that the electric force due to patch potentials of this type contributes
only 0.23\% and 0.008\% of the Casimir force at separations $d=62\,$nm
(the closest separation in this experiment) and 100\,nm, respectively.
These results are based on the theoretical expressions of Ref.~\cite{18},
and the determination of the maximum and minimum sizes
of grains in gold layers covering the test bodies using the atomic force
microscopy images of the surfaces of the plate and sphere. 
With respect to the electric force $F_{\rm el}$ due to the applied
potential $V=0.2\,$V, the patch effect contributes only 0.064\% and 0.0011\%
at separations 62\,nm and 100\,nm, respectively (in this experiment the 
contact potential was determined to be $V_c=3\pm 3\,$mV).
According to
the analysis of Ref.~\cite{10}, at the shortest separations,
$d=160$ and 170\,nm in the experiment using a micromechanical oscillator,
patch potentials contribute only 0.037\% and 0.027\% of the Casimir pressure.
With respect to the electric pressure $P_{\rm el}$ due to
$V-V_c=0.2\,$V, here the patch effect contributes 0.19\% and 0.13\%
at $d=160$ and 170\,nm, respectively. 
There is another type of patch potential due to scratches, adsorbates,
chemical contaminants and dust on the
surface which depends on the applied voltage and, thus, 
significantly influences  the calibration measurements making $V_c$ 
separation-dependent. 
It is generally recognized that such poor quality samples should not be 
used in precision experiments on the Casimir force.
 Thus,  it is unlikely that patch charges are responsible for the
anomalous distance dependence of the gradient of the
electric force observed in Ref.~\cite{Onofrio}.

Here, we present a realistic explanation for the observation of 
Ref.~\cite{Onofrio} that the power of the distance
in the gradient of the electric force  differs from $-2$. 
A key point to note is that Ref.~\cite{Onofrio} used very large 
spheres of radius more than 3\,cm, which approached as close as 20--30\,nm to
the plate. In such a situation the proximity force approximation in the
form (\ref{eq1}) is not valid. To see this we note that Eq.~(\ref{eq1})
was derived for a perfect spherical lens with a constant curvature
radius $R$ at each point of the surface. Reference \cite{Onofrio}
mentions the deviations from ideal spherical geometry and its possible
role at the smallest distances, but considers this only in connection with
the surface roughness. Using the measured rms values of roughness from 1
to 2\,nm, the authors find the respective corrections negligible.
In reality, however, surfaces of large lenses are far from
perfect, even excluding the rms roughness from consideration.
In particular, the typical surface quality of 
centimeter-size surfaces is usually characterized
in terms of the scratch/dig optical surface specification data. 
This means that depending on the quality of lens used, bubbles
or pits with a maximal diameter varying from $30\,\mu$m to 1.2\,mm
are allowed on the surface. There may also be scratches on the surface with
a width varying from 3 to $120\,\mu$m \cite{19}. Surface accuracy is
characterized by the power and irregularity, where power defines the 
deviation of the fabricated surface radius from the radius of a test
surface.
When the separation distance between the sphere and the plate is
sufficiently large, the deviations from perfect spherical shape can
be neglected. Only in this case is the global curvature radius $R$  
important. At short separations, however, local radii of curvature,
which may differ from the global radius by several orders of magnitude
due to the mechanical polishing and grinding of glass lens,
contribute significantly to the result. 

Based on the  above information, we present in Fig.~1 a model of a spherical
lens of radius $R$  containing a region $AB$ of a larger 
curvature radius $R_{AB}=1.6R=49.4\,$mm and a spherical bubble of
$R_{CD}=30\,\mu$m radius. We emphasize that
the height of the sector $AB$ is $H=250\,$nm and the height of the sector
$CD$ is $h=8\,$nm. The imperfections in the large spherical surface, as shown
(not to scale!) in Fig.~1, are well below the error in the determination
of the lens radius $\Delta R=0.15\,$mm. Thus, for a perfect sphere of
radius $R$ the sector $AB$ would have $\tilde{H}=400\,$nm height.
This means that the maximum flattening of the spherical surface in the
region $AB$ is only 150\,nm, i.e., 0.1\% of the allowed error
$\Delta R$ in the radius $R$.

The application of the proximity force approximation to the configuration
in Fig.~1 at small separations results in the modified
coefficient
\begin{equation}
k_{\rm el}^{\rm mod}(d)=\frac{\epsilon_0}{4\pi m_{\rm eff}}\left[
\frac{R_{CD}}{d^2}+\frac{R_{AB}-R_{CD}}{(d+h)^2}-
\frac{R_{AB}-R}{(d+h+H)^2}\right].
\label{eq4}
\end{equation}
\noindent
Numerically, $k_{\rm el}^{\rm mod}(d_0)=k_{\rm el}(d_0)$ at $d_0=30\,$nm.
This equation means that the gradient of the electric force depends
on the separation distance in a far different way than in Eq.~(\ref{eq1}).
As an illustration, in Fig.~2(a) we plot the normalized coefficients 
$k_{\rm el}$, as given by Eq.~(\ref{eq3}) (solid line 1), and 
$k_{\rm el}^{\rm mod}$, as given by Eq.~(\ref{eq4}) (solid line 2), 
as functions of
separation. The normalization factor is equal to 
$N_0\equiv\epsilon_0/(4\pi m_{\rm eff})\times 10^{13}$. 
It can be seen that there is a 
significant deviation between the coefficients obtained for a
perfect spherical lens and that
for the surface shown in Fig.~1. To describe this 
deviation quantitatively,  in Fig.~2(b) we plot the same lines 1 and 2
in a double logarithmic scale. In the same figure the dashed line
shows the dependence of $\tilde{k}_{\rm el}/N_0$ on separation in
accordance with
\begin{equation}
\tilde{k}_{\rm el}(d)=
\frac{\epsilon_0R}{4\pi m_{\rm eff}d_0^{0.3}d^{1.7}}.
\label{eq5}
\end{equation}
\noindent
This expression having a power $-1.7$ instead of $-2$, is shown in
Ref.~\cite{Onofrio} to be consistent with the experimental data
of the measurements of the electric force between a large lens and a plate 
at small separation distances. As is seen in Fig.~2(b), the experimentally
consistent dependence (\ref{eq5}) is well reproduced by the solid line 2
obtained using the standard electric force gradient taking into account
local deviations from a perfect spherical shape, as presented
in Fig.~1.

We emphasize that Fig.~1 shows only one crude model of possible
deviations from sphericity specific for large spherical surfaces.
In precision measurements one should carefully investigate the interaction
region of the large spherical surface microscopically and compute the electric
force numerically by solving Poisson's equation (as done in
Ref.~\cite{16}). Importantly, such complicated problems do not arise
with spheres of small radius. Specifically, the surfaces of polystyrene
spheres of about $100\,\mu$m radius made from liquid phase
are extremely smooth due to surface tension. The investigation of the surface
quality
of such spheres in the scanning electron microscope
did not reveal any scratches or bubbles. However, the same investigation 
has shown 
the presence of bubbles in some $300\,\mu$m and larger 
polystyrene spheres.

In precision electrostatic calibrations, as a part of experiments on
measuring the Casimir force, the following rule is helpful. Depending
on the size and quality
of a spherical body, the minimal separation distance should
be chosen in such a way that the contact potential $V_c$ 
and other basic quantities determined from calibration do not depend
on separation where the calibration procedure is performed.
As an example, in Fig.~3 we present previously unpublished calibration
data for $V_c$ in the experiment on the indirect dynamic determination of
the Casimir pressure between two parallel plates by means of a sphere
oscillating above a micromechanical torsional oscillator \cite{14}.
In this experiment, a sapphire sphere of $R=151.3\pm 0.2\,\mu$m 
radius was used 
and the measurements of the Casimir pressure were performed over the
separation range from 162\,nm to 746\,nm. In Fig.~3 the calibration
results for $V_c$ obtained at 500 different separation distances
ranging from 160.4 to 5150.1\,nm are shown as dots as a function of 
separation. It is seen that the results do not depend on 
separation over a wide separation region including the entire measurement
range of the Casimir pressure. This confirms that proportions between the
sphere radius and the minimum separation are determined correctly.
The resulting mean contact potential is $V_c=15.29\pm 0.13\,$mV.

One more important requirement to precision measurements of the
Casimir force is that the piezo creep and drift should be calibrated
and subtracted. In contrast to experiment \cite{5}, where
continuous voltages were applied to the piezo which was interferometrically
calibrated, or to experiments \cite{10,14}, where the piezo was monitored
interferometrically with a feedback, Ref.~\cite{Onofrio} applies to the 
piezo only static voltages and takes 8--10 minutes to make a measurement.
Then the creep is measured at some large voltage and is scaled linearly 
for the measurement time. This procedure may lead to errors because the
piezo drift is nonlinear with the applied voltage, which might be critical 
at short separation distances.

One can conclude that  contrary to what is claimed in 
Ref.~\cite{Onofrio} the observed ``anomalies'' are irrelevant to the
precision experiments on measuring the Casimir force \cite{2,3,6,7}
mentioned in \cite{Onofrio} and all other performed experiments previously
using the sphere-plate configuration \cite{4,5,8,9,10,11,12,13,14,15,16}.
The experimental precision of these experiments and the measure of
agreement of the obtained results with theory remain the same as was
stated in the original publications after taking account subsequently
published corrections, improvements and reanalyses using more rigorous
statistical approaches (for example, the experimental data of 
Ref.~\cite{5} were later reanalyzed in Ref.~\cite{20}).
It should be mentioned that Ref.~\cite{Onofrio} incorrectly ascribes
the claimed accuracy from 0.1\% to 5\% to the experiments \cite{2,3,6,7}.
In fact the claimed accuracy of these experiments ranges from 1\% to 5\%.
Presently the most precise 
determination of the Casimir pressure using a
micromechanical oscillator is characterized by an experimental error
of 0.2\% and by a 1.9\% measure of agreement between experiment and
theory at the shortest separation of 162\,nm \cite{14}. This experiment,
however, is not mentioned in Ref.~\cite{Onofrio}.

The above remarks demonstrate that the ``anomalous behavior'' of the
electrostatic signal observed in Ref.~\cite{Onofrio} has a clear 
explanation in the mistaken assumption of a perfect spherical shape for 
a mechanically polished and ground large glass lens at nanoscale distances
from a plate.

R.S.D.~acknowledges NSF support through Grants No.~CCF-0508239
and PHY-0701636, and from the Nanoscale Imaging Center at IUPUI.
E.F. was supported in part by DOE under Grant No.~DE-76ER071428.
U.M., G.L.K. and V.M.M. were 
 supported by the NSF Grant No.~PHY0653657 
(computations of the electric force) and
DOE Grant No.~DE-FG02-04ER46131 (precise procedures for piezo
calibrations).
G.L.K. and V.M.M. were also partially supported by
Deutsche Forschungsgemeinschaft, Grant No.~436\,RUS\,113/789/0--4.


\begin{figure}
\vspace*{-4cm}
\centerline{
\includegraphics{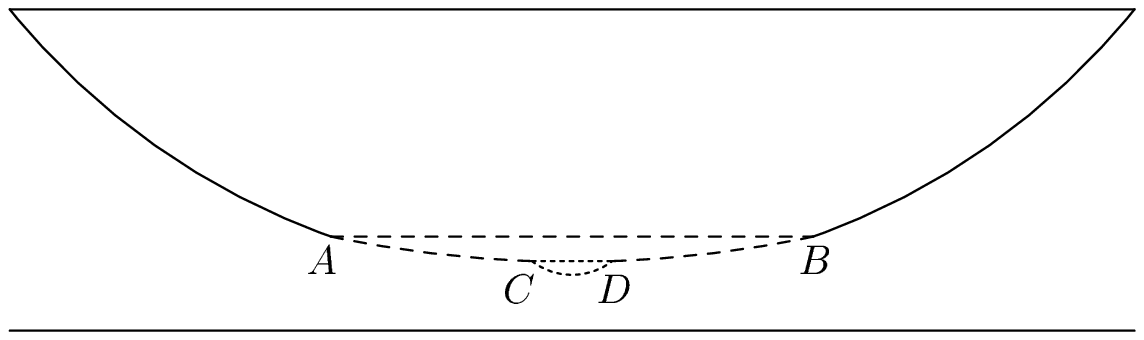}
}
\vspace*{-12cm}
\caption{Model of the surface of the spherical lens of radius $R$ with
local deviations from perfect shape (see text for detail).
Figure is not to scale.
}
\end{figure}
\begin{figure}
\vspace*{0cm}
\centerline{$\phantom{~~~~~~~~~~~~~~~~~~~~~~}$
\includegraphics{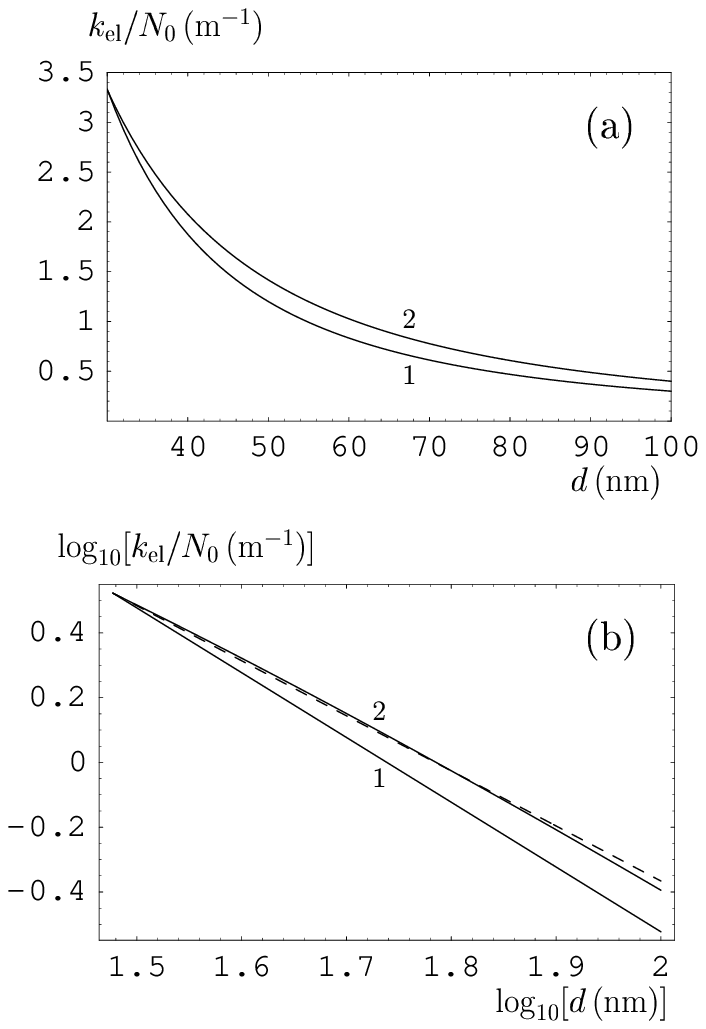}
}
\vspace*{-16cm}
\caption{The normalized coefficient $k_{\rm el}$ in (a) natural and
(b) double logarithimic scales as function of separations. Solid lines
1 and 2 indicate $k_{\rm el}$ and $k_{\rm el}^{\rm mod}$ for a perfect
sphere and for a sphere with local deviations from perfect sphericity.
The dashed line demonstrates $\tilde{k}_{\rm el}$ decreasing as
$d^{-1.7}$.}
\end{figure}
\begin{figure}
\vspace*{-12cm}
\centerline{$\phantom{~~~~~~~~~~~~~~~~~~~~~~}$
\includegraphics{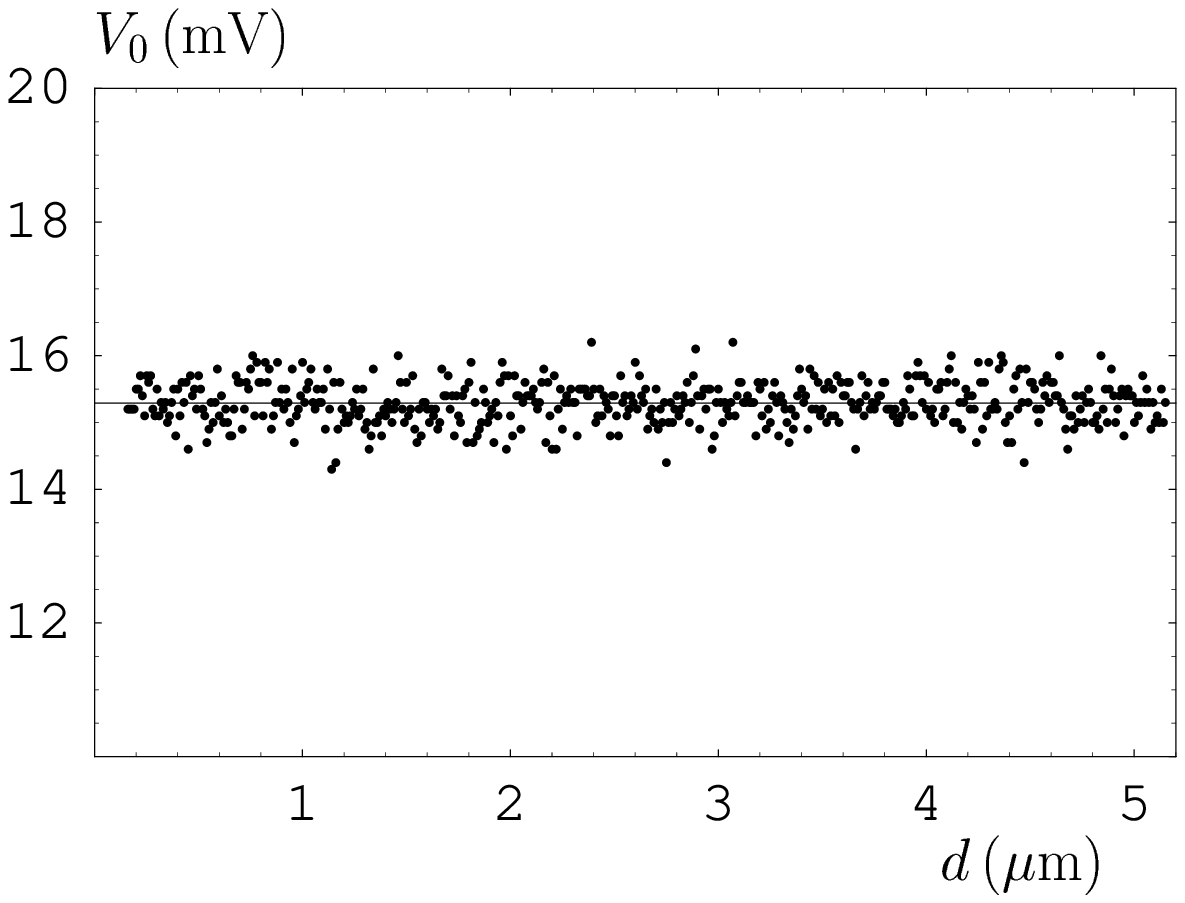}
}
\vspace*{-6cm}
\caption{Calibration results for the contact potential $V_c$ at different
separations are shown as dots. The solid line indicates  the mean value
$V_c=15.29\pm 0.13\,$mV.}
\end{figure}

\end{document}